\documentclass[%
 reprint,
 superscriptaddress,
 amsmath,amssymb,
 aps,
 prl
]{revtex4-2}

\usepackage{float}
\usepackage{xcolor}
\usepackage{graphicx}
\usepackage{dcolumn}
\usepackage{bm}
\usepackage[normalem]{ulem}

\usepackage{tabularx}
\usepackage{verbatim}

\DeclareRobustCommand{\uvec}[1]{{%
	\ifcsname uvec#1\endcsname
		\csname uvec#1\endcsname
	\else
		\bm{\hat{\mathbf{#1}}}%
	\fi
}}

\newcommand{\diff}{\mathop{}\!\mathrm{d}}

\begin{document}

\title{Anomalous sub-kelvin thermal frequency shifts of ultra narrow-linewidth solid state emitters\\}

\author{X. Lin}
\author{M. T. Hartman}
\author{B. Pointard}
\author{R. Le Targat}
\affiliation{%
 LNE-SYRTE, Observatoire de Paris, Universit\' e PSL, CNRS, Sorbonne Universit\' e, 75014 Paris, France\\
}
\author{P. Goldner}
\affiliation{%
Chimie ParisTech, Universit\' e PSL, CNRS, Institut de Recherche de Chimie Paris, 75005 Paris, France\\
}
\author{S. Seidelin}
\affiliation{%
Univ. Grenoble Alpes, CNRS, Grenoble INP and Institut N\' eel, 38000 Grenoble, France\\
}
\author{B. Fang}
\author{Y. Le Coq}
 \email{yann.lecoq@obspm.fr}
\affiliation{%
 LNE-SYRTE, Observatoire de Paris, Universit\' e PSL, CNRS, Sorbonne Universit\' e, 75014 Paris, France\\
}

\date{\today}

\begin{abstract}
We investigate the frequency response of narrow spectral holes in a doped crystal structure as a function of temperature below 1\,K. We identify a particular regime in which this response significantly deviates from the expected two-phonon Raman scattering theory. Namely, near 290\,mK, we observed a behaviour exhibiting a temperature-dependent frequency shift of zero, to first-order. This is of particular interest for applications which require high frequency-stability, such as laser frequency stabilization, as by operating the scheme at this specific point would result in the spectral hole frequency being highly immune to temperature fluctuations, providing the potential for a laser fractional frequency instability as low as $\mathrm{\sim6\times10^{-22}}$ at 1\,s.
\end{abstract}

\maketitle

As the requirements in terms of coherence time and frequency stability in present-day quantum systems are growing continuously, thermal perturbations from the environment are playing an increasingly important role. That is why systems initially operated at room temperature such as ion traps \cite{cryoiontraps} or even optical tweezers \cite{cryotweezers} for quantum computing are now also being implemented in 4\,K cryostats. Moreover, some systems initially operated at 4\,K have shown the potential to greatly benefit from an even lower temperature. One of these systems includes rare-earth doped crystals, which have been studied and exploited in a variety of applications, such as quantum information processing schemes and quantum memories  \cite{Hedges2010,Bussieres2014,Walther2015,Maring2017,Zhong2017,Laplane2017}, quantum optomechanics \cite{molmer2016dispersive,seidelin2019rapid,Ohta2021}, and frequency stabilization schemes \cite{thorpe2011frequency,leibrandt2013absolute,galland2020double,cook2015laser}. Whereas many of these studies are indeed possible at standard liquid helium-4 temperatures, several of these applications might greatly benefit from a sub-kelvin environment \cite{Askarani2021}. More precisely, in the case of rare-earth dopants, the frequency shift of spectral features, according to the two-phonon Raman scattering theory, is supposed to vary with the temperature as $T^{4}$, with a zero frequency displacement starting from zero kelvin \cite{konz2003temperature}. Thus, in principle, at low temperatures the center-frequency will be less perturbed by temperature fluctuations. 

Eliminating frequency fluctuations is particularly important in the context of laser stabilization schemes, where a laser is locked on a narrow spectral feature \cite{thorpe2011frequency,leibrandt2013absolute,galland2020double,cook2015laser}. In this context, an additional technical factor further motivating the lower temperatures is that the achievable absolute temperature instability is significantly lower at lower temperatures. Furthermore, in our study at sub-kelvin temperatures, our findings indicate an improvement beyond the $T^{4}$ expectation, where we discover an anomaly around 290\,mK that proves valuable for further eliminating the effect of temperature noise on the transition frequency.\\

In this work we investigate the sub-kelvin (100 to 900\,mK) behavior of a system consisting of europium ions ($\rm Eu^{3+}$) doped into a $\rm Y_2SiO_5$ crystal (Eu:YSO), chosen due to its attractive coherence properties \cite{equall1994ultraslow,oswald2018characteristics}. The europium ions can substitute an yttrium atom in two different nonequivalent sites in the YSO matrix, referred to as site 1 and site 2 (vacuum wavelengths of 580.04\,nm and 580.21\,nm, respectively \cite{thorpe2011frequency}). We employ the techniques of spectral hole burning \cite{sellin_programmable_1999} in order to achieve deep spectral holes with linewidths below 3\,kHz (in full width at half maxima) within an otherwise inhomogeneously broadened (typically 2\,GHz at our 0.1 at.\% europium doping density) absorption spectrum. Our laser system (with a fractional frequency instability of $\mathrm{\sim2\times10^{-15}}$ at 1\,s, as measured using an independent optical frequency comb and Fabry-Perot-referenced laser system) and data acquisition setup are described in great detail in previous articles \cite{galland2020mechanical,galland2020double, lin2023multi-mode}. We highlight only the main differences here. Previously, we have studied the temperature behavior using a standard 4\,K cryogenic system \cite{Zhang2023}. Here, to reach sub-kelvin temperatures, we use a dilution refrigerator system, that combines compressing and pumping of a $\mathrm{^3He-^4He}$ mixture through elements at various temperatures and fluids conductance to produce a significant cooling power below the 3.5\,K temperature limit of our pulse-tube based system. This dilution stage is a 40\,cm high prototype add-on to our commercial MyCryoFirm Optidry system, which now provides approximately $30~\mu$W of cooling power at 100\,mK, and allows cooling the 1\,cm$^3$ crystal from 3.5\,K to 100\,mK in 2\,h. Its temperature is measured using a composite sensor combining a conventional $100~\Omega$ resistor and a ruthenium oxide sensor that is calibrated against a SI-traceable sensor.  The temperature is measured using a Lakeshore 350 controller which is also used for temperature stabilization by applying feedback on a heating resistor near the $\mathrm{^3He-^4He}$ copper mixing-chamber. The lowest achievable free-running temperature is around 80\,mK. An active stabilization of the temperature (acting at time constants $> 1\,\mathrm{s}$) can thus be obtained in the 100-900\,mK range (above which the dilution system can no longer efficiently cool against overheating of the mixing chamber), or at temperatures higher than 3.5\,K by turning off the dilution stage and using only the pulse-tube cooler.

A computer, connected to the cryo controller via ethernet link, sets and retrieves the temperature of the cryostat. Using silver lacquer for thermal contact, the Eu:YSO crystal is mounted on a copper plate, which itself is tightly clamped to the base plate of the $\mathrm{^3He-^4He}$ mixing chamber whose temperature is actively controlled. To verify that the use of the silver lacquer does not perturb the measurements (due to potential strain in the crystal arising from differences in thermal expansion coefficients of copper, the silver lacquer and the crystal itself), we perform an initial series of measurements at temperatures above 3.5\,K which can, contrary to the sub-kelvin measurements, be compared with existing literature \cite{thorpe2011frequency,thorpe2013shifts}. This is done by burning a spectral hole in the inhomogeneous profile of the crystal using a laser with a fixed frequency, at 3.5\,K, and subsequently scanning a probe laser across the structure to measure its frequency position at varying temperature values from 3.5 to 9\,K and noting its temperature dependent frequency shift, $f_\mathrm{shift}$. The uncertainty on the value of the scanned frequency is $\mathrm{\sim100\,Hz}$, mainly limited by the processing in the frequency control system. 

\begin{figure}[t]
    \includegraphics[width=80mm]{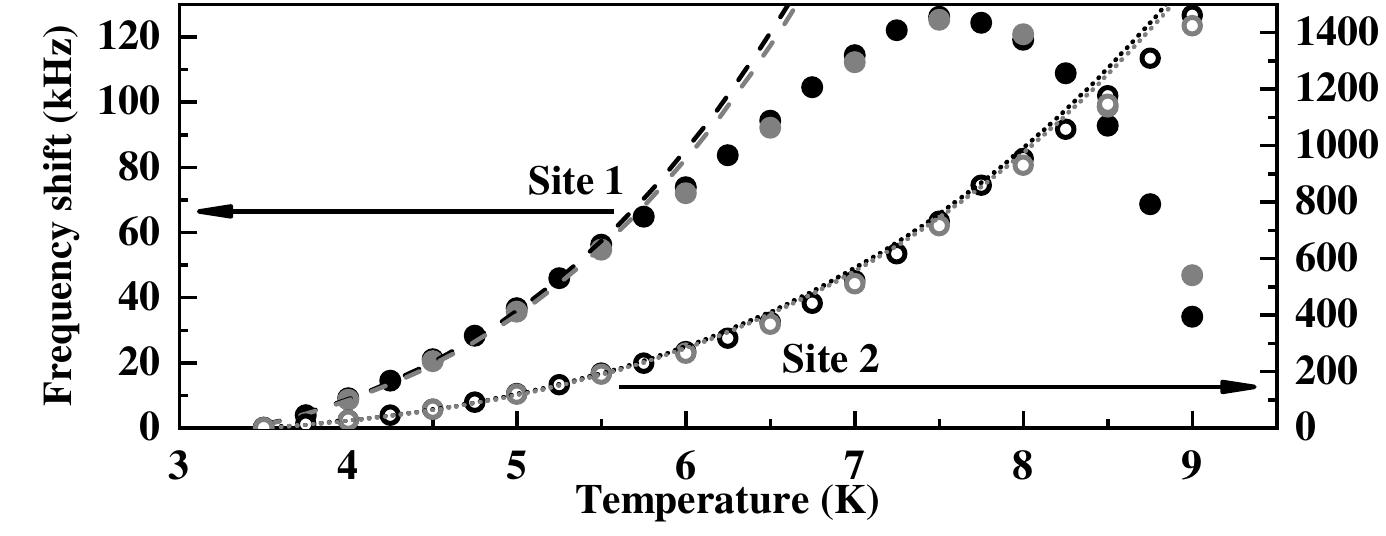}
    \caption{\label{fig:freq_shift_3_K} Frequency shifts as a function of temperature measured above 3\,K, for crystal site 1 (a) and crystal site 2 (b), and for two different orientations of the crystal (triangle markers for D1 vertical, and squares markers for D1 horizontal). The black and gray dashed lines correspond to a $\alpha T^{4}$ curve fit for D1 vertical and horizontal, respectively.}
\end{figure}

The crystal is, throughout this work, optically probed along the b crystalline axis with the polarization along the D1 dielectric axis, \emph{i.e.} the configuration that maximizes absorption \cite{Ferrier2016}. To verify that strain due to differential thermal expansion does not impact the experiment, the measurement is done initially with the crystal D1-axis mounted vertically, and then repeated after turning the crystal 90 degrees, re-lacquering, and re-mounting, such that the D1-axis is then horizontal. More precisely, according to our previous work \cite{galland2020mechanical}, the frequency shift as a function of stress applied along the D1 and the D2 axes are of opposite sign. A rotation should therefore modify the direction of any potential stress-induced frequency-shift, and thus lead to different results, should a mounting-induced strain have a significant effect.

\begin{figure*}[t]
    \includegraphics[width=\textwidth]{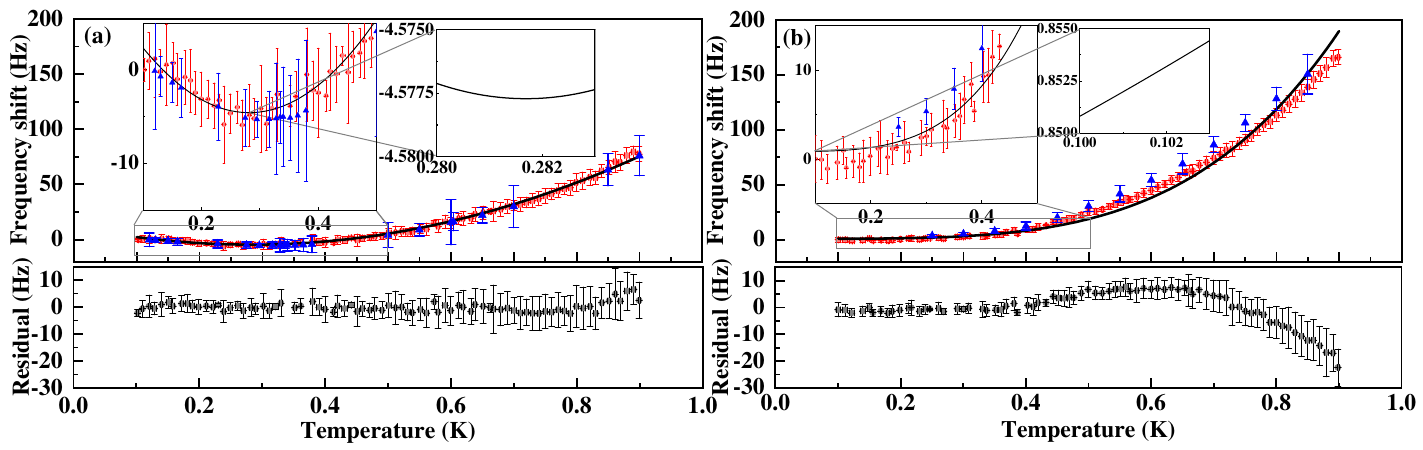}
    \caption{\label{fig:freq_shift_1_K} Top: Measured frequency shifts and curve fits in the interval from 0.1 to 0.9\,K, for crystal site 1 (a) and 2 (b), in both cases measured with the D1 axis oriented vertically (blue corresponds to the modulation method, and red the ramp method, see text). Bottom: the residuals from the curve fit. The error bars correspond to the statistical standard deviations.}
\end{figure*}

The results for temperatures above 3\,K, at both spectroscopic sites using both mounting orientations, are plotted in FIG. \ref{fig:freq_shift_3_K}. We fit this data in the 3.5 to 5.5\,K range (also utilized by previous work \cite{thorpe2013shifts}) to a $f_\mathrm{shift}=\alpha T^{4}+\mathrm{cte}$ model as predicted by a two-phonon Raman scattering process \cite{konz2003temperature}. The fit results of these spectral scan measurements are summarized in the ``$>3\,\mathrm{K}$'' column of TABLE \ref{tab:fitsmK}. Although the difference of measurement is slightly larger than the stated statistical uncertainty, they still agree within 2 percent, which may be considered as the systematic uncertainty of our measurement. For comparison, in TABLE \ref{tab:fitsmK} we also include previous results obtained by a group at NIST, who used a suspended Eu:YSO crystal \cite{thorpe2013shifts}, in good agreement with our results. 

The downward trend of the curve starting at around 7.5\,K for site 1 (FIG. \ref{fig:freq_shift_3_K}(a)) is also in accordance with what has been previously observed at NIST, further supporting the fact that thermally induced strain on the crystal is negligible. While intriguing as a point of temperature insensitivity, the local maximum occurring around 7.5\,K is unsuitable for laser stabilization as, at such high temperature, the spectral holes broaden past $\mathrm{FWHM}>10\,\mathrm{kHz}$ and experience a drastic decrease in contrast, diminishing the frequency discrimination of the spectral feature. Furthermore, at these temperatures the spectral holes have lifetimes of order 100 minutes \cite{konz2003temperature}. The zero first-order sensitivity point can, however, be artificially engineered to be around 4\,K, by applying a specific background gas pressure, which requires a buffer gas and a high degree of control of its pressure \cite{Zhang2023,thorpe2011frequency}. This approach, however, presents significant technical challenges for realizing a leak-proof helium gas enclosure around the crystal and, more importantly, may be sensitive to differences of temperature equalization time constants, as it relies on the crystal, the helium background gas and the mechanical enclosure to be at thermal equilibrium. 

At dilution temperatures, we need to measure frequency shifts that are significantly smaller than those measured above 3\,K.  Here, the 100\,Hz resolution obtained in a simple frequency scan is no longer sufficient. We therefore employ a different approach, where we frequency lock the probe laser to a spectral hole pattern, with a low-noise high-precision technique described in our recent publication \cite{lin2023multi-mode}. When modifying the temperature set-point of the crystal, we compare (via an optical frequency comb) the deviations in the frequency of the spectral-hole locked-laser against a separate, conventional Fabry-Perot-based ultra stable laser which exhibits a fractional frequency stability below $10^{-15}$ at 1\,s and increasing to $8\times 10^{-15}$ at $\approx$ 80\,s. We can therefore measure accurately the changes in the position of the spectral hole pattern with an resolution of a few Hz (depending on time constant), provided they occur in this range of duration.

Two distinct strategies were utilized for changing the temperature in a sufficiently rapid but controlled way. 

In the first strategy, we apply a lock-in detection scheme. To this end, we set the crystal temperature to a value $T_{0}$ and then modulate the set-point of the cryo-controller with a sinusoidal function.  In this measurement we simultaneously record the time series of three signals: the registered set-point (nominally a discretized sinusoid), the actual measured temperature, and the measured frequency of the spectral-hole locked laser as beat against the comb+Fabry-Perot referenced laser. Careful synchronization of these records is based on a local network time protocal (NTP) server-client method for the time-tagging temperature set-points and measurement records, and pulse per second (PPS)-based method for frequency comb measurements (with linear interpolation for one-to-one correspondence). Both NTP and PPS signals are referenced to UTC(OP). In post-processing, the set-point sinusoid is used to demodulate the measured temperature signal and optical frequency signal both in-phase and in-quadrature. The measurement was repeated at several measurement temperatures, $T_0$. The modulation frequency dependant dephasing between these measurement records reveals a constant pure delay component (related to instrumental synchronisation) and a single pole (within measurement accuracy and time constant range) low-pass filter related to temperature evolution dynamics in the crystal+mount+mixing chamber ensemble. The selected modulation frequency, $f_\mathrm{mod}=31.6\,\mathrm{mHz}$, was chosen as to be below this measured characteristic thermal frequency, $f_\mathrm{th}\approx70\,\mathrm{mHz}$. The modulation depth, $m=0.01\textup{--}0.09$ was selected based on the available cooling power at $T_0$. The number of modulation periods (between $30\textup{--}100$) was selected to achieve a signal to noise ratio $>10$. The ratio of the amplitude of frequency modulation to the amplitude of measured temperature modulation gives the temperature sensitivity,$\frac{\diff{f_\mathrm{shift}}}{\diff{T}}$, of the spectral hole frequency at each measurement point. In order to be easily compared with the results of the second measurement strategy (described next), integrating this quantity demonstrates the frequency shift of the spectral hole versus temperature, plotted in FIG. \ref{fig:freq_shift_1_K} (blue markers). 

The second strategy is to apply a rapid linear upward ramp of temperature set-points (update approximately every 1\,s) for a total duration of $\sim 80$ seconds. Although fast enough to benefit in full of our Fabry-Perot based optical frequency reference stability, it is too fast for the temperature servo to follow the set-point accurately. As a consequence, the correspondence between the temperature and the optical frequency measurements is retrieved from synchronized temperature readout. Note that in these time constants, it is only possible to apply the temperature ramp upward, as it is much easier for the temperature servo to warm-up than cool down (which is inherently limited by cooling power). A small correction is applied to temperature measurements to take into account the low-pass filter dynamics previously mentioned (this correction doesn't modify the overall behavior, and only changes the fitted parameters described below by about 5 percent. Several repetitions of the process provides data shown in FIG. \ref{fig:freq_shift_1_K} (red markers), where the data points are horizontally binned on 10\,mK intervals, and the vertical error bars are measured statistical deviations on these bins.

\begin{table*}[t]
\begin{centering}
\caption{\label{tab:fitsmK} Table of fitted coefficients for 100mK-900m\,K data ($aT^2+bT+c$ fit for site 1 and $\alpha T^4+c$ for site 2, where $a$, $b$, $c$, $\alpha$ are free parameters) and data above 3\,K. The values of $c$ are not shown as they have no physical meaning, being determined by the burning frequency; VD1 and HD1: vertical and horizontal D1 axis, respectively; M. T.: magic temperature, the location of the local minimum.}
\begin{tabularx}{17.8 cm}{>{\centering\arraybackslash}X>{\centering\arraybackslash}X>{\centering\arraybackslash}X>{\centering\arraybackslash}X>{\centering\arraybackslash}X>{\centering\arraybackslash}X>{\centering\arraybackslash}X>{\centering\arraybackslash}X}
\hline
\hline
$ $
& 
& 
& \multicolumn{4}{c}{$\mathrm{<1\,K}$}
& $\mathrm{>3\,K}$\\
\cline{4-7}
$\mathrm{Site}$
& $\mathrm{Orientation}$
& $\mathrm{Method}$
& $\mathrm{\alpha\,(Hz/K^4)}$ 
& $\mathrm{a\,(Hz/K^2)}$ 
& $\mathrm{b\,(Hz/K)}$
& $\mathrm{M.\,T.\,(mK)}$
& $\mathrm{\alpha\,(Hz/K^4)}$
\\
\hline
$\mathrm{1}$ 
& $\mathrm{VD1}$  
& $\mathrm{Modulation}$
& $-$ 
& $198(10)$ 
& $-115(3)$
& $290(26)$
& $69(5)$
\\
$\mathrm{1}$ 
& $\mathrm{VD1}$  
& $\mathrm{Ramp}$ 
& $-$ 
& $209(3)$ 
& $-118(3)$
& $282(12)$
& $-$
\\
$\mathrm{1}$ 
& $\mathrm{HD1}$  
& $\mathrm{Ramp}$ 
& $-$ 
& $234(11)$ 
& $-141(8)$
& $301(30)$
& $-$
\\
$\mathrm{1}$ 
& $\mathrm{VD1}$  
& $\mathrm{Scan}$ 
& $-$ 
& $-$ 
& $-$
& $-$
& $74(1)$
\\
$\mathrm{1}$ 
& $\mathrm{HD1}$  
& $\mathrm{Scan}$ 
& $-$ 
& $-$ 
& $-$
& $-$
& $71(2)$
\\
$\mathrm{1}$ 
& $-$  
& $\mathrm{NIST\,Results}$\cite{thorpe2013shifts} 
& $-$ 
& $-$ 
& $-$
& $-$
& $76(15)$
\\
\hline
$\mathrm{2}$ 
& $\mathrm{VD1}$ 
& $\mathrm{Modulation}$ 
& $260(18)$ 
& $-$ 
& $-$
& $-$
& $262(7)$
\\
$\mathrm{2}$ 
& $\mathrm{VD1}$  
& $\mathrm{Ramp}$
& $288(4)$ 
& $-$ 
& $-$
& $-$
& $-$
\\
$\mathrm{2}$ 
& $\mathrm{HD1}$  
& $\mathrm{Ramp}$ 
& $274(5)$ 
& $-$ 
& $-$
& $-$
& $-$
\\
$\mathrm{2}$ 
& $\mathrm{VD1}$  
& $\mathrm{Scan}$
& $-$ 
& $-$ 
& $-$
& $-$
& $251(1)$
\\
$\mathrm{2}$ 
& $\mathrm{HD1}$  
& $\mathrm{Scan}$ 
& $-$ 
& $-$ 
& $-$
& $-$
& $247(1)$
\\
$\mathrm{2}$ 
& $-$  
& $\mathrm{NIST\,Results}$\cite{thorpe2013shifts}
& $-$ 
& $-$ 
& $-$
& $-$
& $250(50)$
\\
\hline
\hline\\
\end{tabularx}
\end{centering}
\end{table*}

The results at spectroscopic site 2, shown in FIG. \ref{fig:freq_shift_1_K}(b), follow approximately the $T^{4}$-curve as predicted by the two-phonon Raman scattering theory. As in the high temperature measurements, we repeat the measurement after turning the crystal 90 degrees (such that D1 axis becomes horizontally oriented) with no visual difference regarding the curve shape (points omitted from FIG. \ref{fig:freq_shift_1_K} for clarity), only the obtained coefficients differ slightly from before, and still roughly agree with the high-temperature measurement. Note that the systematic trend observed in fit residuals may, however, hint at supplementary physical processes not taken into account in the two-phonon Raman scattering model.

\begin{figure}[b]
    \includegraphics[width=\linewidth]{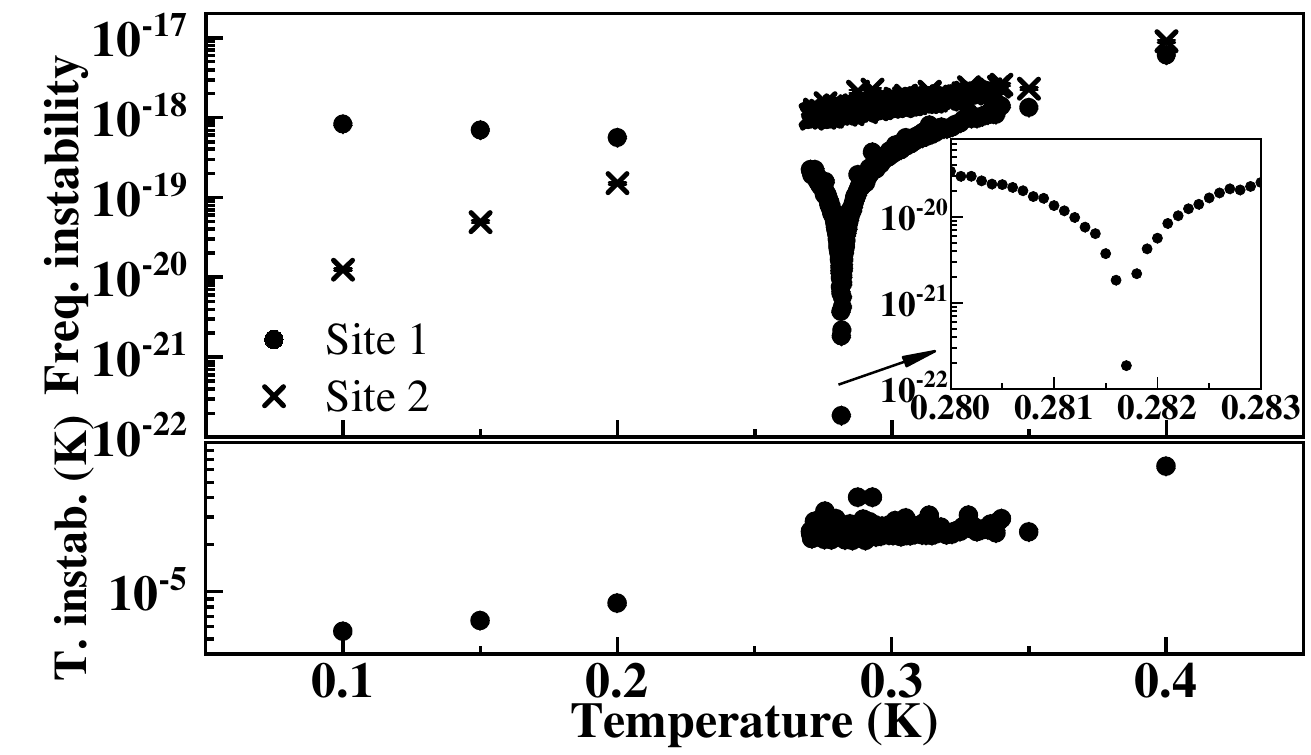}
    \caption{\label{fig:temp_stability} The measured temperature instabilities (bottom) and the projected corresponding temperature induced fractional frequency instabilities at 1\,s (top) at different temperatures. The D1 axis is vertical. Top: The square ($\blacksquare$) and triangle ($\blacktriangle$) markers are the projected fractional frequency instabilities of site 1 and 2 at different temperatures, respectively. The solid and dashed lines are the projected fractional frequency instabilities from 0.1 to 0.4 K of site 1 and 2, respectively. The inset shows the details around the zero-frequency-drift temperature point. Bottom: The dots ($\cdot$) are the crystal temperature instabilities at different temperatures. }
\end{figure}

Plotted in FIG. \ref{fig:freq_shift_1_K}(a), the case of site 1, on the other hand, demonstrates a behavior that largely deviates from the $T^4$ prediction. Here, the curve exhibits a clear local minimum around 290\,mK, for which the frequency is insensitive to temperature fluctuations at first-order.  Without a theoretical model conveying the observed behavior, we use an effective second order polynomial in order to fit the data.  The fit results are summarized in ``$<1\,\mathrm{K}$'' column of TABLE \ref{tab:fitsmK}, which includes ``magic temperature'' (M.T.), the temperature location of the local minimum.

In order to assess the corresponding achievable frequency stability of a laser locked to a spectral hole, we first measure the Allan deviation of temperature at 1\,s (where temperature servo effect is minimal) in 100-700\,mK range. Here, for our cryostat, we noted lower temperature deviations at lower operating temperatures, shown in FIG. \ref{fig:temp_stability} (bottom).  According to the measured temperature stability near 290\,mK (at 1\,s) and the fitting result in TABLE \ref{tab:fitsmK}, the temperature-induced fractional instability at the zero-drift point would be $\approx6\times10^{-22}$ at 1\,s, orders of magnitudes below the limit currently predicted from other sources of instability or requirement of any current optical clock for quantum projection noise (or even Heisenberg) limited operation. The maximum rejection is, however, only obtained at a sharply defined temperature for a specific sample.  Undoubtedly, re-characterization would be required for other manufactured samples and measurement of its exact position is necessary for each particular crystal setup. 

An alternative strategy would involve reducing the temperature even further, below the 100\,mK value. However, in our case, as discussed, we have been able to use a simple, relatively inexpensive and small footprint dilution stage inserted directly into our existing 4-Kelvin pulse tube device.  For temperatures below 100\,mK, a significantly larger dilution refrigerator apparatus is required, increasing cost and complexity, thus motivating the approach outlined above.\\

In conclusion, we have experimentally demonstrated the existence of a temperature near 290 mK for which a laser frequency stabilized to a spectral hole in site 1 of Eu:YSO is, to first order, insensitive to temperature variations of the crystal.  We evaluated its potential for making temperature fluctuations a virtually non-contributing factor for metrology applications.  Due to the lack of symmetry of the $\rm Y_2SiO_5$ crystal, a theoretical model capable of mimicking the anomalous behaviour for site 1 in the neighbourhood of 290\,mK is far from straightforward to elaborate. A similar behaviour, also for site 1, has been previously observed around 7.5\,K \cite{Zhang2023,thorpe2011frequency}, as well as around 32\,K \cite{konz2003temperature}. The physical origin behind these anomalies is not yet understood, but it has been speculated that it involves structural instabilities of the crystal lattice \cite{konz2003temperature} and this work provides further measurements that might help shedding light on the underlying physical mechanisms.\\

We acknowledge financial support from Ville de Paris Emergence Program, the Région Ile de France DIM C’nano and SIRTEQ, the LABEX Cluster of Excellence FIRST-TF (ANR-10-LABX-48-01) within the Program “Investissement d’Avenir” operated by the French National Research Agency (ANR), the 15SIB03 OC18 and 20FUN08 NEXTLASERS projects from the EMPIR program cofinanced by the Participating States and from the European Union’s Horizon 2020 research and innovation program, and the UltraStabLaserViaSHB (GAP-101068547) from Marie Skłodowska-Curie Actions (HORIZON-TMA-MSCA-PF-EF) from the European Commission Horizon Europe Framework Programme (HORIZON).


%

\end{document}